# Unit Cell-Level Thickness Control of Single-Crystalline Zinc Oxide Nanosheets Enabled by Electrical Double Layer Confinement

Xin Yin, Yeqi Shi, Yanbing Wei, Yongho Joo, Padma Gopalan, Izabela Szlufarska, Xudong Wang*

Department of Material Science and Engineering, University of Wisconsin – Madison

Email: xudong.wang@wisc.edu

**Abstract**

Ionic layer epitaxy (ILE) has recently been developed as an effective strategy to synthesize nanometer-thick 2D materials with a non-layered crystal structure, such as ZnO. The packing density of the amphiphilic monolayer is believed to be a key parameter that controls the nanosheet nucleation and growth. In this work, we systematically investigated the growth behavior of single-crystalline ZnO nanosheets templated at the water-air interface by an anionic oleylsulfate monolayer with different packing densities. The thicknesses of ZnO nanosheets were tuned from one unit cell to four unit cells, and exhibited good correlation with the width of $Zn^{2+}$ ion concentration zone (the Stern layer) underneath the ionized surfactant monolayer. Further analysis of the nanosheet sizes and density revealed that the nanosheet growth was dominated by the steric hindrance from the surfactant monolayer at lower surface pressure; while the nucleation density became the dominating factor at higher surface pressure. The ZnO nanosheets exhibited a decreasing work function as the thickness reduced to a few unit cells. This research validated a critical hypothesis that the nanosheet growth is self-limited by the formation of a double layer of ionic precursors. This work will open up a new way towards controlled synthesis of novel 2D nanosheets from non-layered materials with a thickness down to one unit cell.

**Introduction**

Two dimensional (2D) nanomaterials, such as graphene and transition metal dichalcogenides (TMDs), have attracted great interests as a new class of materials with remarkable chemical, physical and mechanical properties associated with their atomic thickness. For example, graphene, the 2D form of carbon, has shown outstanding room temperature electron mobility,[1] very high thermal conductivity,[2] large photon absorption,[3] and extraordinary mechanical strength.[4] TMDs, such as $MoS_2$, $WS_2$, exhibit thickness-related tunable electronic bandgaps, which enabled enormous application potentials for transistors,[5] photodetectors,[6] and electroluminescent devices.[7] One common feature shared among all these 2D nanomaterials is their layered crystal structures. The weak van der Waals interaction between the strongly-bonded atomic layers makes the synthesis of these 2D atomic sheets accessible, either from top-down methods, like exfoliation by micromechanical cleavage,[8] ionic intercalation in solution,[9] and ultrasonication;[10] or from bottom-up methods, such as chemical vapor deposition.[11-12] Nevertheless, many functional materials do not possess such a 2D-favorable crystal structure. Instead, they exhibit strong chemical bonding along all crystallographic directions, presenting a grand challenge in synthesizing 2D atomic sheets.

Recently, a new bottom-up technique named ionic layer epitaxy (ILE) was developed to grow nanometer-thick 2D materials at the water-air interface, which opened numerous opportunities in creating a new family of 2D nanomaterials from non-layer crystals.[13-14] In ILE, amphiphilic molecules (*e.g.*, surfactants) self-assemble into a monolayer at the water-air interface, serving as a template to direct the nucleation and growth of a crystalline nanosheet underneath. Single-crystal ZnO nanosheet with ~2 nm thickness has been shown as a representative example of this unique 2D growth strategy. However, the growth mechanism was not completely clear for understanding the growth process. It was hypothesized that the ionized head groups of the surfactants stabilize an electrical double layer in their vicinity and that the formation of this double layer controls the growth. Specifically, it was proposed that the concentrated zone (the Stern layer) provides a highly localized precursor supersaturation region self-limiting the nanosheet thickness. This hypothesis, if validated, will bring an unprecedented control of nanosheet thickness down to the sub-nanometer level. Here, using the ZnO nanosheet growth system, we systematically investigated the correlation between the packing density of the oleylsulfate monolayer and the thickness of single-crystalline ZnO nanosheets. The $Zn^{2+}$ ion concentration profile under the surfactant monolayer was calculated by molecular dynamics (MD) simulation. The enhanced charge density from more densely packed surfactants induced a wider $Zn^{2+}$ ion concentration zone, which yielded thicker nanosheets. Through this strategy, the average nanosheet thickness was controlled from one to four unit cells. The influences of surfactant packing density to the nanosheet size and density were also investigated, generating new insights into the ILE growth kinetics. As the nanosheet thickness reduced from four unit cells to one unit cell, its work function monotonically decreased from 4.91±0.03 eV to 4.73±0.02 eV, showing a potential for designing heterojunctions with tunable band alignment. These results validate the previously hypothesized model of electrical double layer self-limiting growth in ILE and provide an effective solution toward creating atomic-thick crystalline nanosheets from non-layered crystals.

**Experimental Section**

**ZnO nanosheet growth.** The surfactant monolayer was formed and compressed in a home-made trough shown in Figure 1A. In a typical synthesis, 60 mL aqueous solution containing 25 mM zinc nitrate ($Zn(NO_3)_2$) and hexamethylenetetramine (HMT) was prepared in a glass vial and immediately transferred into the trough. 20 μL Chloroform solution of sodium oleyl sulfate (~0.1 vol %) was dispersed on the solution surface. Twenty minutes was allowed for equilibrium of the monolayer and evaporation of the chloroform. Subsequently, the trough was capped by a lid to form a closed reaction environment and placed in a 60 °C convection oven for 1 hour and 45 minutes to harvest ZnO nanosheets, which could be then scooped using an arbitrary substrate for further characterization.

**Characterization.** Zeiss LEO 1530 Schottky-type field-emission scanning electron microscope was used to study the morphologies of the samples. FEI TF30 transmission electron microscope operated at 300 kV was used to study the crystal structure of the samples. EDX in the STEM mode on a probe aberration corrected FEI Titan at 200 kV. Atomic force microscopy (AFM) tomography images were obtained using XE-70 Park Systems. The surface pressure–bar position isotherm of the monolayer was measured by a Langmuir–Blodgett trough (KSV NIMA Medium size KN 2002) with a Wilhelmy balance (Platinum plate).

**Simulation Methods.** Simulations are performed with GROMACS 5.1.2 software package and GROMOS 54A7 force field (Supporting information S1.3).

**Results and Discussion**

ZnO nanosheets were synthesized by ILE in a home-made trough reactor (Figure S1) directed by an anionic oleylsulfate monolayer at the water-air interface (see supplementary materials for the synthesis details). The schematic ILE growth under controlled surfactant packing density is shown in Figure 1A. The movable crossing bar enabled a fine control of the dispersion surface area with a resolution of 0.11 mm$^2$ per step. Here, the surface pressure was measured as a function of the bar position by a Langmuir–Blodgett trough with a Wilhelmy balance (Platinum plate) to represent the packing density of the surfactant monolayer when a 20 μL of 0.1 vol% surfactant solution in chloroform was added to the surface of the Zn precursor aqueous solution (Fig. 1B).[15] This measurement revealed that the surfactant packing density can be continuously tuned within a relatively large range following an almost linear relation. Thus, ZnO nanosheets were synthesized *via* the standard ILE procedure at five different surface pressures (23.09 mN/m, 16.38 mN/m, 10.29 mN/m, 5.84 mN/m, and 3.09 mN/m) to investigate the surfactant packing density influences. After 105-minute growth at 60 °C, the ZnO nanosheets were transferred onto a Si wafer from the water surface. Typical nanosheet morphologies obtained at the five different surface pressures are shown in Figure 1C. At the surface pressure of 23.09 mN/m, no nanosheets but only arbitrarily shaped nanoparticles were observed (point I). When the surface pressure decreased to 16.38 mN/m, well-developed triangular nanosheets with sizes of ~10 microns were formed in large quantities (point II). ZnO nanosheets could also be obtained when the surface pressure further decreased to 10.29 mN/m, 5.84 mN/m, and 3.09mN/m, but these sheets appeared to have different sizes and quantities per unit surface are (points III-V). Further reduction of the surface pressure (controlled by increasing the reaction surface area in the trough) yielded no growth of ZnO at the interface. This initial observation revealed that there existed a surface pressure window, within which the nanosheet morphology could evolve. Too low or too high surfactant packing density would lead to no growth or particle formation, respectively.

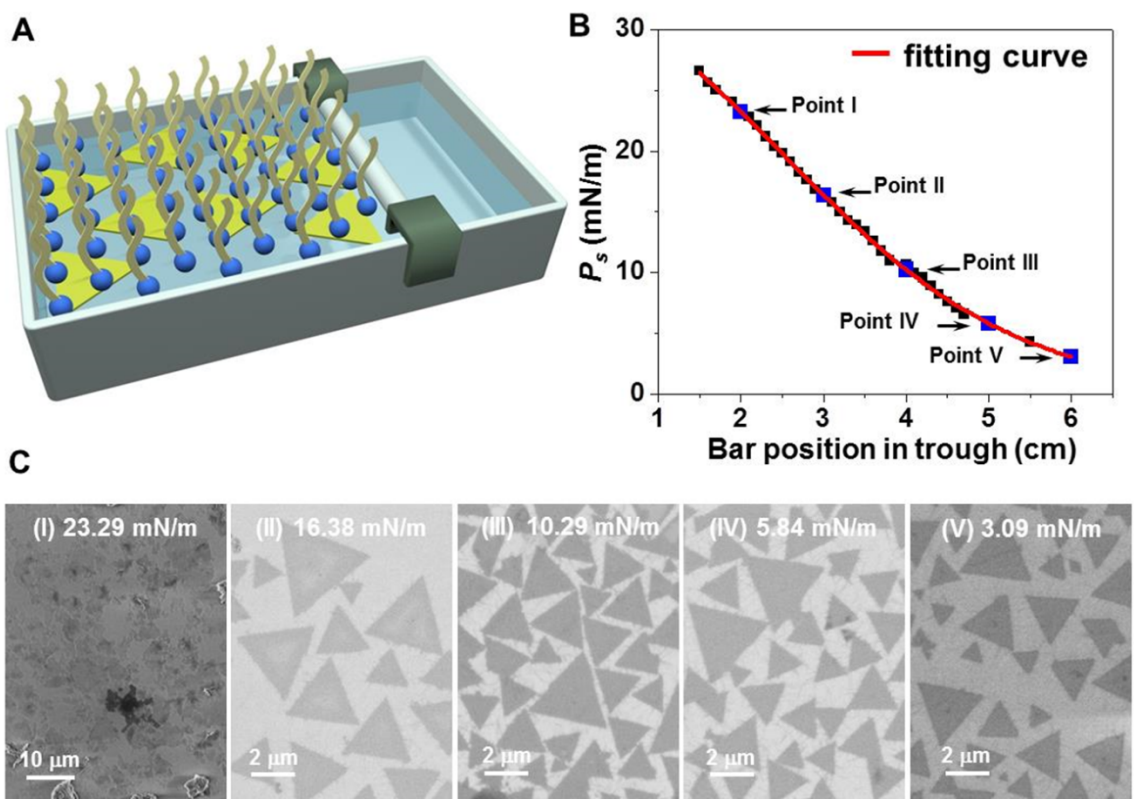

**Figure 1. Nanosheet growth under controlled surfactant packing density. (A)** Schematic setup of growing nanosheets in the trough with a movable bar to control the packing density of the surfactant monolayer. **(B)** Surface pressure measured at room temperature as the bar position change. **(C)** SEM images of the nanosheet growth results at the surface pressures of 23.29 mN/m (I), 16.38 mN/m (II), 10.29 mN/m (III), 5.84 mN/m (IV) and 3.09 mN/m (V). Scale bars are 10 μm.

The nanosheets synthesized in the trough were then characterized to confirm their chemical composition and crystal structure. Figure 2A shows a typical SEM image of triangular ZnO nanosheets with sizes up to 20 μm as-transferred onto a Si substrate. They were synthesized at the surface pressure of 16.38 mN/m. The regular triangular shape with three 60° corners was consistent with other synthesis results under all different surface pressures. The AFM topography image in Figure 2B displays the sharp side edges and smooth top surface of the nanosheet. The thickness was measured to be ~2.35 nm. The low magnification TEM image in Figure 2C shows a nanosheet (with brighter contrast) rested on a holey carbon TEM grid. The wrinkles of the nanosheet resulted from the transferring process could also be clearly observed, indicating the excellent flexibility of the nanosheet. A high resolution TEM (HRTEM) image from the nanosheet shown in Figure 2C revealed the high-quality single-crystalline lattice with a typical hexagonal arrangement (Figure 2D). Corresponding fast Fourier transfer (FFT) pattern (bottom inset) matched well with the selected area electron diffraction pattern (SAED, top inset), confirming the wurtzite crystal structure of the ZnO nanosheet with the surface to be the (0001) plane and the three equivalent $\{10\bar{1}0\}$ planes as the side edges. EDS mapping at the corner area marked by the blue dashed box in Figure 2C showed a uniform distribution of oxygen and zinc elements in the nanosheet (Figure 2E and 2F). The line profiles of oxygen and zinc elements (from the green dashed line in Figure 2C) also exhibited a good match across the entire nanosheet (Figure 2G), confirming their nearly 1:1 atomic ratio and good stoichiometry of the ZnO composition (Table S1). These characterizations confirmed that the triangular nanosheets synthesized by ILE in the trough were wurtzite ZnO single crystals.

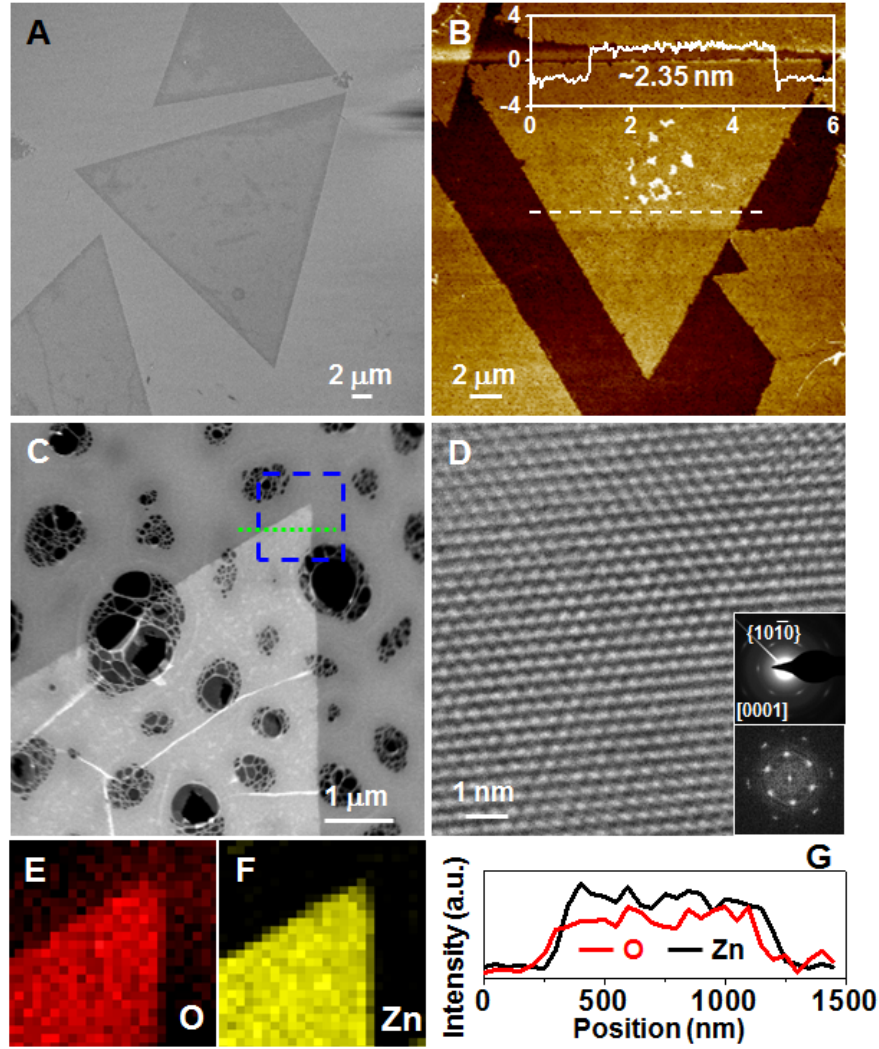

**Figure 2. Crystal structure and composition characterization of the nanosheets.** **(A)** SEM image of triangular ZnO nanosheet as-transferred to a Si substrate surface. **(B)** Topography AFM image showing the nanometer thickness and excellent surface flatness of the nanosheets. Inset is the height profile along the white dashed line revealing the thickness is 2.35 nm. **(C)** Low magnification TEM image of a ZnO nanosheet rest on a TEM grid. **(D)** High resolution TEM image showing the perfect crystal lattice of Wurtzite ZnO without any observable defects. Fast Fourier transform pattern (lower inset) and selective area electron diffraction pattern (upper inset) match well and both confirmed the hexagonal crystal structure. Elemental mapping of **(E)** oxygen and **(F)** zinc in the area marked by the blue dashed box in **(C)**. **(G)** O and Zn element intensity profile along the green dashed line in **(C)**.

In order to obtain a better understanding of the influence of the surfactant monolayer on the nanosheet growth at different surface pressures, MD simulations were performed to model the surfactant monolayer at the water/vacuum interface (simulation details are included in supplementary materials S1.3). The evolution of the oleysulfate molecules with increasing surface pressure (as controlled by the density of surfactants) is shown in Figure 3A. The range of pressures in the simulations is consistent with the pressures under which the nanosheets were experimentally achievable (Figure 1C). The corresponding simulated distribution of $Zn^{+2}$ resulting from the increasing electric field strengths due to packing of the surfactant monolayer is plotted in Figure 3B, where the red line represents the interface, separating the vacuum and the aqueous solution. The zinc concentration curves show a peak near the interface, right beneath the surfactant monolayer. This peak indicates the presence of a thin two-dimensional Zn-concentrated zone (less than 2 nm thick) along the interface due to the electrostatic attraction from the negatively charged monolayer.

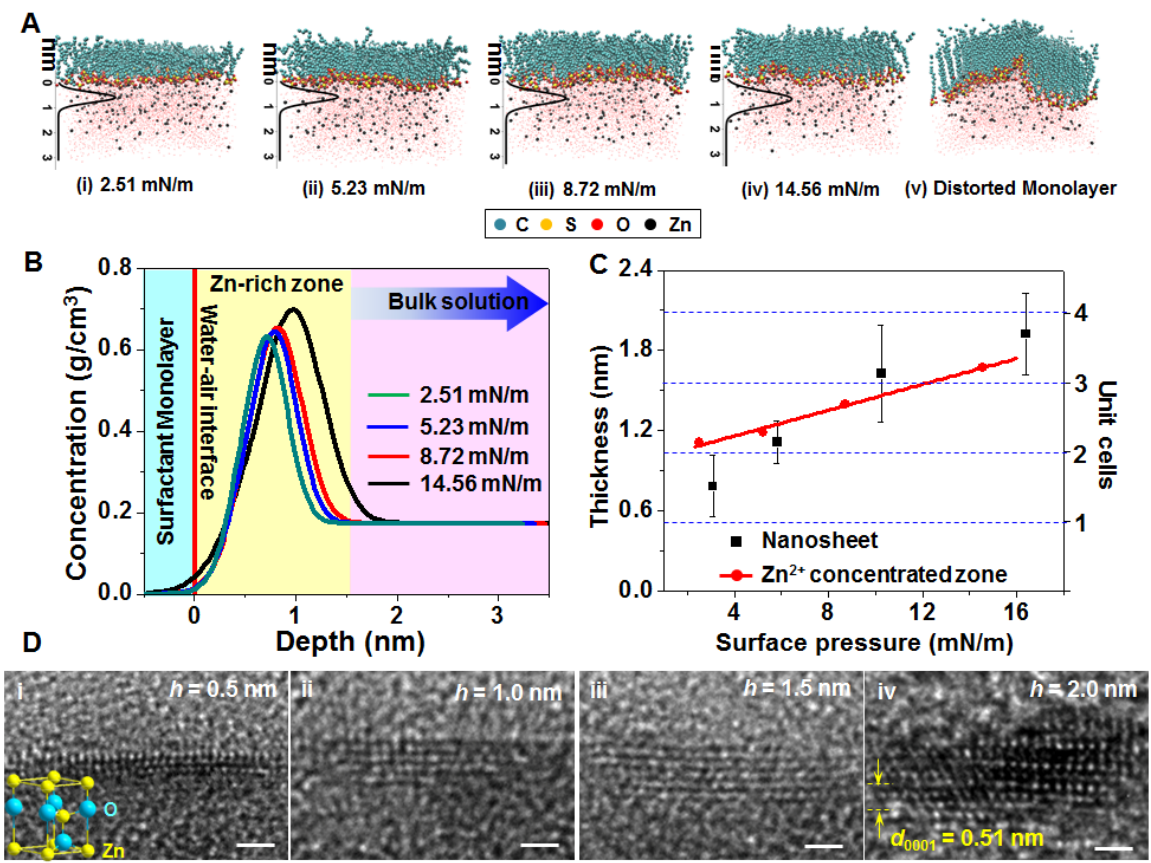

**Figure 3. Influence of the surfactant monolayer packing density on the nanosheet thickness.** **(A)** MD simulation generated $Zn^{2+}$ ion distribution under the surfactant monolayer at different

packing density. From (i) to (iv), the surface pressures are 2.51 mN/m, 5.23 mN/m, 8.72 mN/m and 14.56 mN/m, respectively. (v) represents the case with higher surface pressure where the surfactant is no longer a flat monolayer. **(B)** $Zn^{2+}$ ion concentration profiles underneath the surfactant monolayer with four different surface pressures. Sky blue represents the surfactant monolayer. Light yellow represents the Zn-concentrated zone (the Stern layer). Lavender represents the bulk solution. **(C)** Plots of the nanosheet thickness (black squares) and the width of Zn-concentrated zone (red dots) as functions of the surface pressure. The numbers of ZnO unit cell are highlighted by dashed blue lines. **(D)** Cross-sectional HRTEM images of ZnO nanosheets with a thickness from one to four unit cells. Inset shows one unit cell of Wurtzite ZnO.

This Zn-concentrated zone was found to evolve with the surface pressure. At low surface pressure of 3.09 mN/m, the area occupied by one molecule was large. As a result, the surfactant molecules were not uniformly ordered, as evidenced by the random packing of the carbon tails shown in Figure 3A(i). This dilute anionic surfactant packing induced a relatively weak electric field due to the low charge density. Consequently, relatively few zinc ions were attracted to the interface leading to a less concentrated and thinner zinc zone. When surface pressure was increased to 5.23 mN/m, the space between molecules became smaller and the molecules became more close-packed (Figure 3A(ii)). Thus, the charge density and therefore the electric field were both increased, resulting in a larger number of zinc ions aggregating along the interface and forming a thicker zinc-concentrated zone. Increasing the surface pressure even further (to 8.72 mN/m) resulted in a uniform ordering of the molecules (approaching the solid phase of the surfactants), an increase in the electric field was, and therefore also a thicker zinc concentrated zone (Figure 3A(iii)). Similarly, at the surface pressure of 14.56 mN/m, all the tails of the surfactant molecules were lined up along the direction perpendicular to the interface due to the smaller space between the molecules (Figure 3A(iv)). Further increase in the surface pressure led to formation of wrinkles and distorted packing (Figure 3A(v)) due to the strong repulsion between molecules (where no nanosheets could be obtained). The above results show that increasing the surface pressure (by condensing surfactant molecules) results in a transition in packing of surfactant molecules at the water/vacuum interface from random, to uniform and ordered, to distorted. The pressure regime where ZnO nanosheets can be grown experimentally corresponds to the regime where surfactants in MD simulations are ordered and form a closely packed and a relatively flat interface. MD simulations also revealed that the evolution of the morphology and density of the surfactants with pressure has a significant impact on the thickness of the Zn-concentrated region. The thickness of this region grows with increasing surface pressure (Figure 3C). We hypothesized that the thickness of the Zn-concentrated layer will affect the thickness of the experimentally grown ZnO nanosheets at corresponding pressures.

To test this hypothesis, the thicknesses of ZnO nanosheets grown at different surface pressures were measured by AFM. Multiple nanosheets (>10) were measured in each sample and their average values were plotted in Figure 3C. Typical AFM images of the ZnO nanosheets from each sample were included in the supplementary materials (Figure S2). It can be clearly seen that as the surface pressure increased, the thickness of the as-grown nanosheets also increased monotonically. More importantly, its trend closely tracked the width of the $Zn^{2+}$ ion concentration zone predicted by MD simulations, providing evidence for a strong correlation between the $Zn^{2+}$ ion Stern layer and the crystalline nanosheet thickness. It should be noted that at such small thicknesses, increase of nanosheet thickness may not be continuous given the length of one ZnO unit cells is 0.52 nm along the [0001] direction. To guide the eyes, four dashed lines were added in Figure 3C to highlight the thickness of different numbers of unit cell. The average film thickness obtained from surface pressure of 5.84 mN/m and 10.29 mN/m (1.12 ± 0.16 nm and 1.63± 0.36 nm, respectively) matched well to the length of two and three unit cells. Nevertheless, the thinnest nanosheets, which were obtained at a surface pressure of 3.09 mN/m, were measured to be 0.78 ± 0.23 nm, slightly larger than one unit cell. It might be because of the relatively larger width of the $Zn^{2+}$ ion Stern layer compared to the one unit cell thickness at this growth condition, and thus the resulted ZnO nanosheets would exhibit a combination of one and two unit cell thicknesses. As the surface pressure reached 16.38 mN/m, the nanosheet thickness rose to 1.93± 0.31 nm. This value was slightly smaller than the length of four unit cells, which might be attributed to the insufficient $Zn^{2+}$ ion concentration to sustain the growth of four complete unit cells.

The unit cell-level thickness control was further verified by cross-sectional HRTEM images of the nanosheets (Figure 3D and S4). Along the c-direction of Wurtzite ZnO, one unit cell has a thickness of 0.52 nm, which corresponds to three layers of bright dots (Zn ions) in the HRTEM image, *i.e.* the (0002) plane as shown in Figure 3D(i). From samples collected from higher surface pressure of 5.84 mN/m, 10.29 mN/m, and 16.38 mN/m, the number of plane (0002) increased to 5, 7 and 9, respectively, corresponding to 2, 3, and 4 unit cells as shown in Figure 3D(ii) – (iv). This analysis supports the claim that by controlling the packing density of the surfactant monolayer one can digitally tune the nanosheet thickness with a resolution of one unit cell all the way down to the thickness of one unit cell.

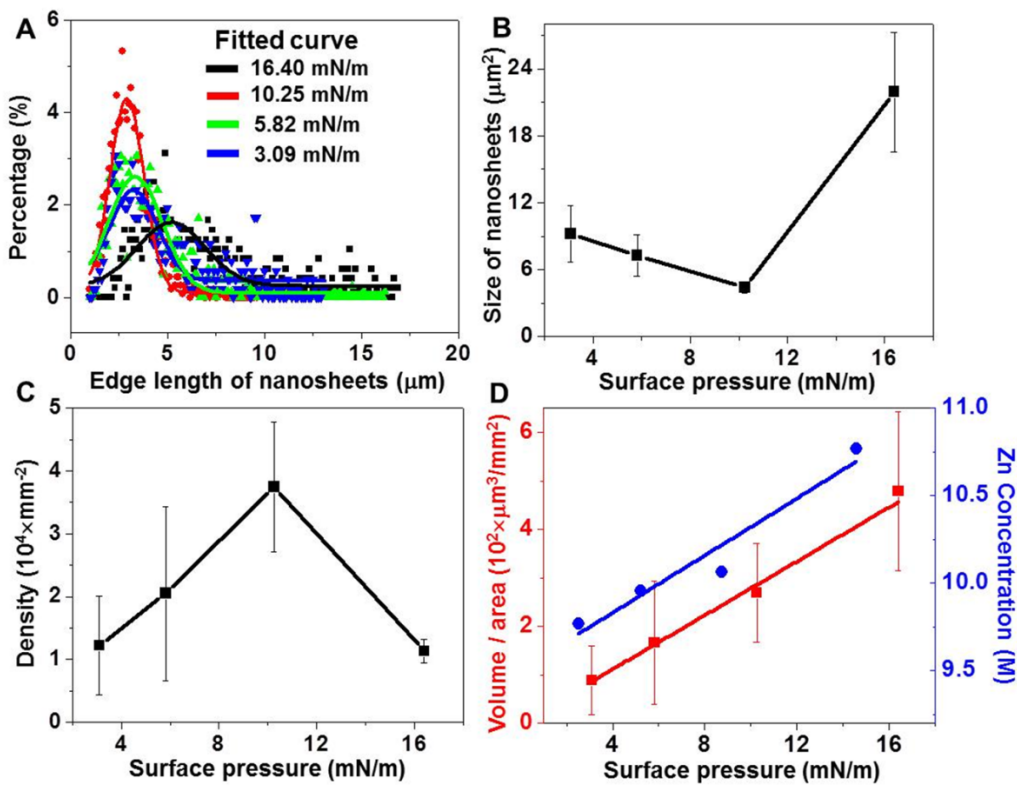

**Figure 4. Growth kinetics investigation. (A)** Size distribution of the nanosheets at four surface pressures. **(B)** Plot of nanosheet size as a function of surface pressure. **(C)** Plot of nanosheet density as a function of surface pressure. **(D)** Plots of ZnO nanosheet coverage on water surface and the calculated $Zn^{2+}$ ion concentration as functions of surface pressure. Both relations follow the same trend confirming the ZnO nanosheets were formed from the $Zn^{2+}$ source only in the Stern layer.

The influences of surfactant packing density were further investigated by statistic analysis on the nanosheet geometry over a large quantity. The size distributions at the four surface pressures, which were represented by the edge length of the triangular nanosheets, are plotted in Figure 4A. For each surface pressure, more than 1,000 nanosheets were measured. All the curves showed the Gaussian distribution with a long tail towards larger sizes. At the surface pressure of 10.25 mN/m, the size distribution was the narrowest with the peak position of 2.91 μm. At the surface pressures of 5.82 mN/m and 3.09 mN/m, the size distributions were slightly broadened and the peak positions slightly increased to 3.30 μm and 3.32 μm, respectively. The largest size distribution was obtained at the surface pressure of 16.40 mN/m, which also yielded the largest peak position of 5.1 μm. From this statistic analysis, the average size of individual nanosheet as a function of surface pressure is summarized in Figure 4B by assuming all the nanosheets are equilateral triangles and there were no overlaps among them. It shows that the average size started to decrease from 9.23 ± 2.55 μm$^2$ to 4.41 ± 0.49 μm$^2$ as the surface pressure increased from 3.09 mN/m to 10.25 mN/m, and then increased to 21.93 ± 5.34 μm$^2$ when the surface pressure further rose to 16.40 mN/m. The smallest nanosheet size was obtained at 10.25 mN/m.

Since it was previously hypothesized that the growth of crystalline nanosheets is self-limited by the available precursor source provided by the Stern layer under the surfactant monolayer,[13] here the number of nanosheets per unit surface area, *i.e.* the nanosheet density, was also analyzed to test this hypothesis. As shown in Figure 4C, the density dependence on the surface pressure exhibited the opposite trend than the size dependence on surface pressure. With the increase of surface pressure, the density increased from $1.22 \times 10^4$ mm$^{-2}$ at 3.09 mN/m and reached the maximum number of $3.74 \times 10^4$ mm$^{-2}$ at 10.25 mN/m. The density decreased back to $1.13 \times 10^4$ mm$^{-2}$ as the surface pressure further increased to 16.40 mN/m. By combining the size, density and thickness relations, the total nanosheet volume per area could be obtained, which provided a direct measure of the amount of Zn$^{2+}$ ions. As shown in Figure 4D, the volume of ZnO nanosheets increased monotonically from 89 μm$^3$/mm$^2$ to 479 μm$^3$/mm$^2$ with the increase of the surface pressures (red curve). It followed the same relation as the Zn$^{2+}$ ion concentration in the Stern layer calculated from MD simulation (blue curve in Figure 4D). This comparison provided a strong support to our hypothesis that the crystalline ZnO nanosheets were formed from the Zn$^{2+}$ source only in the Stern layer. Within the nanosheet growth window, more Zn$^{2+}$ ions in the Stern layer will produce more crystalline ZnO nanosheets.

In addition, the opposite size and density relations shown in Figure 4B and 4C provide valuable insights into the ILE growth kinetics. Based on the observation of time-dependent nanosheet evolution, we previously proposed that single crystalline nanosheets were formed *via* oriented attachment of numerous nano-crystallites nucleated from the Stern layer.[13-14] It follows that, the nanosheet formation is controlled by two competing factors: nucleation density and surfactant density. Higher nucleation density facilitates the attachment of discrete crystallites during the growth, which is favorable for the formation of large nanosheet. Thermodynamically, the nucleation density is directly related to the $Zn^{2+}$ ion concentration in the Stern layer, which is proportional to the surfactant packing density. At the same time, higher surfactant packing density induces larger steric hindrance to the oriented attachment, and thus limits the formation of large nanosheet. Our experiments showed that at lower surface pressure from 3.09 mN/m to 10.25 mN/m, the steric hindrance from the surfactant monolayer dominated the nanosheet growth. That is, the higher the surfactant packing density, more resistance would be experienced during the oriented attachment of the crystallites, resulting in the decreased nanosheet size and increased nanosheet density. As the surface pressure further increased from 10.25 mN/m to 16.40 mN/m, the change of the surfactant packing density became less significant due to the already very tight space between the surfactant molecules. Nevertheless, the higher $Zn^{2+}$ ion concentration in the Stern layer provided a much lower nucleation energy barrier to boost the nucleation density, as thus largely promoted the oriented attachment of crystallites. Therefore, much larger nanosheets with a low density were obtained in this region.

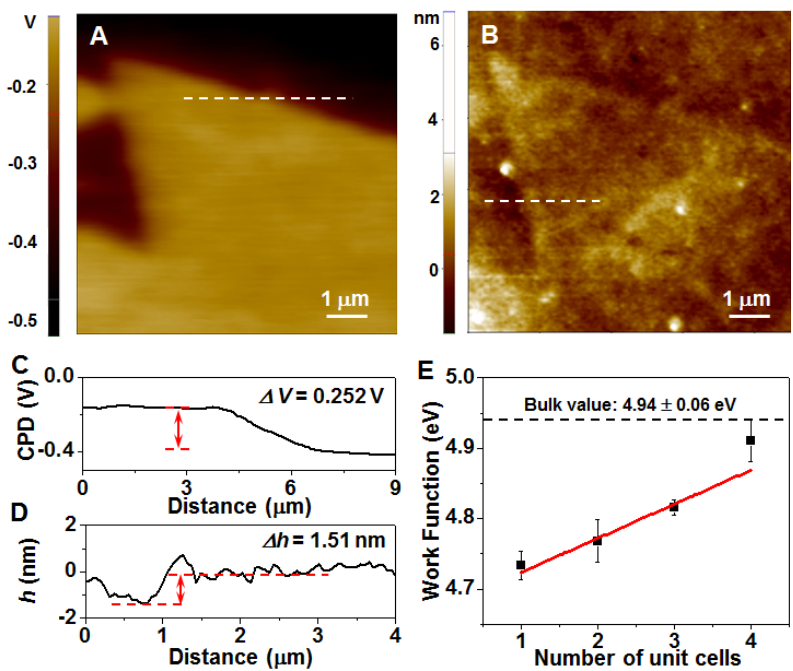

**Figure 5. Work function investigation. (A)** EFM image of a ZnO nanosheet rested on gold surface. **(B)** Corresponding AFM topographic image of the ZnO nanosheet shown in **A**. **(C, D)**

Line profiles of voltage and height data extracted from **A and B**, respectively. **(E)** Work function measured by EFM as a function of nanosheet thickness. The bulk value was marked by the dashed line.

Work function is a fundamental electronic property of semiconductor materials. Here, the thickness-dependent work function was analyzed by measuring the surface potentials using scanning Kelvin probe microscopy (SKPM), where the work function was extracted from surface potential differences. The nanosheets with different thickness were supported by a Si substrate coated with a 100-nm-thick Au thin film as the reference. The surface potential mapping of one representative sample is shown in Figure 5A, revealing the uniform surface potential distribution over the entire nanosheet. The corresponding topography image is shown in Figure 5B. Along the dashed line in Figure 5A and B, the surface potential difference between ZnO nanosheet and Au thin film was measured to be 0.252 V (Figure 5C), while the nanosheet had a thickness of 1.51 nm, i.e. 3 unit cells (Figure 5D). The work function of the nanosheets was then calculated from the surface potentials following the equation $\Phi_{ZnO} = \Phi_{Au} - e \times (CPD_{Au} - CPD_{ZnO})$,[16] where $\Phi_{ZnO}$ is the work function of ZnO nanosheet, $\Phi_{Au}$ is the work function of Au, which is 5.1 eV, $CPD_{Au}$ and $CPD_{ZnO}$ are contact potential difference between the tip and Au film, and the ZnO nanosheet, respectively. It should be noted that while the work function of gold in the range of 4.74 eV to 5.54 eV have been reported, 5.1 eV is the often used value, particularly for flat surfaces.[17-18] Thus, the work function evolution with the thickness increasing from one unit cell to four unit cells were extracted and plotted in Figure 5E, which shows increasing trend from 4.73±0.02 eV to 4.91±0.03 eV, approaching the bulk value of 4.94 ± 0.06 eV (Supplementary Materials Figure S5-S8). This thickness-dependent work function provides a good flexibility in designing heterojunctions with tunable band alignment.

**Conclusion**

In summary, the influences of surfactant monolayer packing density were systematically investigated in ILE ZnO nanosheet growth. By adjusting the surfactant spreading area at the water surface, nanometer-thick single crystalline ZnO nanosheets were synthesized with various sizes, thicknesses and densities. MD simulations were performed to reveal the $Zn^{2+}$ ion distribution profile underneath the surfactant monolayer at different packing densities (*i.e.*

surface pressure). It was found that both the $Zn^{2+}$ concentration and the width of the $Zn^{2+}$ concentrated zone (the Stern layer) increased monotonically with increasing surfactant packing density. Comparing experimental measurements with the simulation results revealed an excellent match between the nanosheet thickness and the Stern layer width, confirming the hypothesis that the thickness of the Stern layer controls the thickness of the grown films. As the surface pressure was adjusted from 3.09 mN/m to 16.40 mN/m, ZnO nanosheets with one to four unit cell thickness were achieved. Further analysis of the nanosheet size and density yielded more insights into the nanosheet growth kinetics. It suggested that nanosheet growth was dominated by the steric hindrance from the surfactant monolayer at lower surface pressure; while the nucleation density became the dominating factor at higher surface pressure. The ZnO nanosheets with reduced thickness exhibited lower work function indicating a potential to achieve tunable band alignment in semiconductor device design. This fundamental study of ZnO nanosheets growth validated a critical hypothesis of the self-limited thickness control in ILE. It further shed light on ILE growth kinetics in correlation to the surfactant packing density. This work will open up a new method to control the synthesis of novel 2D nanosheets from non-layered materials with thickness down on one unit cell.

**Supporting Information**

Synthesis process, characterization methods, simulation methods, image of the trough, EDX characterization, 3D topography AFM images, concentration profiles of $H_2O$, headgroup, tail and $Zn^{2+}$ in the solution, cross sectional SEM and TEM, and details of surface potential measurement. These materials are available free of charge via the Internet at http://pubs.acs.org.

**Acknowledgement:** This work is supported by Army Research Office (ARO) under grant W911NF-16-1-0198 and National Science Foundation through the University of Wisconsin Materials Research Science and Engineering Center (DMR-1121288).

**References:**

1. Mayorov, A. S.; Gorbachev, R. V.; Morozov, S. V.; Britnell, L.; Jalil, R.; Ponomarenko, L. A.; Blake, P.; Novoselov, K. S.; Watanabe, K.; Taniguchi, T., Micrometer-scale ballistic transport in encapsulated graphene at room temperature. *Nano Lett.* **2011,** *11*, 2396-2399.

2. Balandin, A. A., Thermal properties of graphene and nanostructured carbon materials. *Nat. Mater.* **2011,** *10*, 569-581.

3. Nair, R. R.; Blake, P.; Grigorenko, A. N.; Novoselov, K. S.; Booth, T. J.; Stauber, T.; Peres, N. M.; Geim, A. K., Fine structure constant defines visual transparency of graphene. *Science* **2008,** *320*, 1308-1308.

4. Lee, C.; Wei, X.; Kysar, J. W.; Hone, J., Measurement of the elastic properties and intrinsic strength of monolayer graphene. *Science* **2008,** *321*, 385-388.

5. Radisavljevic, B.; Radenovic, A.; Brivio, J.; Giacometti, i. V.; Kis, A., Single-layer MoS2 transistors. *Nature nanotechnology* **2011,** *6*, 147-150.

6. Lopez-Sanchez, O.; Lembke, D.; Kayci, M.; Radenovic, A.; Kis, A., Ultrasensitive photodetectors based on monolayer $MoS_2$. *Nat. Nanotechnol.* **2013,** *8*, 497-501.

7. Sundaram, R.; Engel, M.; Lombardo, A.; Krupke, R.; Ferrari, A.; Avouris, P.; Steiner, M., Electroluminescence in single layer MoS2. *Nano Lett.* **2013,** *13*, 1416-1421.

8. Novoselov, K.; Jiang, D.; Schedin, F.; Booth, T.; Khotkevich, V.; Morozov, S.; Geim, A., Two-dimensional atomic crystals. *Proc. Natl. Acad. Sci. U. S. A.* **2005,** *102*, 10451-10453.

9. Ida, S.; Shiga, D.; Koinuma, M.; Matsumoto, Y., Synthesis of hexagonal nickel hydroxide nanosheets by exfoliation of layered nickel hydroxide intercalated with dodecyl sulfate ions. *J. Am. Chem. Soc.* **2008,** *130*, 14038-14039.

10. Choucair, M.; Thordarson, P.; Stride, J. A., Gram-scale production of graphene based on solvothermal synthesis and sonication. *Nat. Nanotechnol.* **2009,** *4*, 30-33.

11. Wei, D.; Liu, Y.; Wang, Y.; Zhang, H.; Huang, L.; Yu, G., Synthesis of N-doped graphene by chemical vapor deposition and its electrical properties. *Nano Lett.* **2009,** *9*, 1752-1758.

12. Strupinski, W.; Grodecki, K.; Wysmolek, A.; Stepniewski, R.; Szkopek, T.; Gaskell, P.; Gruneis, A.; Haberer, D.; Bozek, R.; Krupka, J., Graphene epitaxy by chemical vapor deposition on SiC. *Nano Lett.* **2011,** *11*, 1786-1791.

13. Wang, F.; Seo, J.-H.; Luo, G.; Starr, M. B.; Li, Z.; Geng, D.; Yin, X.; Wang, S.; Fraser, D. G.; Morgan, D., Nanometre-thick single-crystalline nanosheets grown at the water-air interface. *Nat. Commun.* **2016,** *7*, 10444.

14. Wang, F.; Yin, X.; Wang, X., Morphological control in the adaptive ionic layer epitaxy of ZnO nanosheets. *Extrem. Mech. Lett.* **2016,** *7*, 64-70.

15. McElhinny, K. M.; Huang, P.; Joo, Y.; Kanimozhi, C.; Lakkham, A.; Sakurai, K.; Evans, P. G.; Gopalan, P., Optically reconfigurable monolayer of azobenzene-donor molecules on oxide surfaces. *Langmuir* **2017**, 2157–2168.

16. Kim, J. H.; Lee, J.; Kim, J. H.; Hwang, C.; Lee, C.; Park, J. Y., Work function variation of MoS2 atomic layers grown with chemical vapor deposition: The effects of thickness and the adsorption of water/oxygen molecules. *Appl. Phys. Lett.* **2015,** *106*, 251606.

17. Giovannetti, G.; Khomyakov, P.; Brocks, G.; Karpan, V. v.; Van den Brink, J.; Kelly, P., Doping graphene with metal contacts. *Phys. Rev. Lett.* **2008,** *101*, 026803.

18. Domanski, A. L.; Sengupta, E.; Bley, K.; Untch, M. B.; Weber, S. A.; Landfester, K.; Weiss, C. K.; Butt, H.-J. r.; Berger, R. d., Kelvin probe force microscopy in nonpolar liquids. *Langmuir* **2012,** *28*, 13892-13899.

**TOC.**

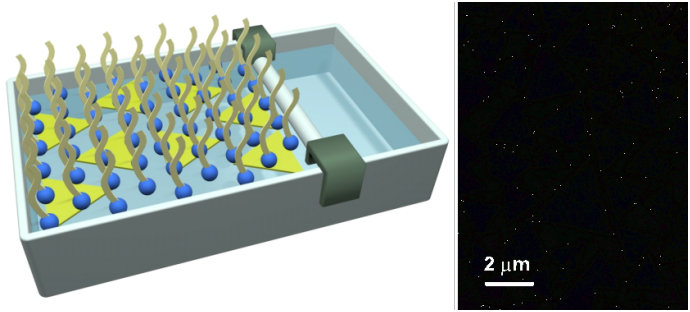